\documentclass[aps,prl,twocolumn,preprintnumbers,superscriptaddress,nofootinbib]{revtex4-1}
\usepackage[utf8]{inputenc}
\usepackage{graphicx}
\usepackage{psfrag}
\usepackage{amssymb}
\usepackage{amsmath}
\usepackage{epstopdf}
\usepackage{color}
\newcommand{\bi}{\begin{itemize}}
\newcommand{\ei}{\end{itemize}}
\newcommand{\bea}{\begin {eqnarray}}
\newcommand{\eea}{\end{eqnarray}}
\newcommand{\be}{\begin{eqnarray}}
\newcommand{\ee}{\end{eqnarray}}
\newcommand{\ket}[1]{|\,#1\,\rangle}          

\newcommand{\LCm}{{\scriptscriptstyle -}} 
\newcommand{\LCp}{{\scriptscriptstyle +}}

\begin{document}
\title{Proton structure from a light-front Hamiltonian}
\author{Chandan Mondal}
\email{mondal@impcas.ac.cn} 
\affiliation{Institute of Modern Physics, Chinese Academy of Sciences, Lanzhou 730000, China}
\affiliation{School of Nuclear Science and Technology, University of Chinese Academy of Sciences, Beijing 100049, China}
\author{Siqi Xu}
\email{xsq123@impcas.ac.cn} 
\affiliation{Institute of Modern Physics, Chinese Academy of Sciences, Lanzhou 730000, China}
\affiliation{School of Nuclear Science and Technology, University of Chinese Academy of Sciences, Beijing 100049, China}
\author{Jiangshan Lan}
\email{jiangshanlan@impcas.ac.cn}
\affiliation{Institute of Modern Physics, Chinese Academy of Sciences, Lanzhou 730000, China}
\affiliation{School of Nuclear Science and Technology, University of Chinese Academy of Sciences, Beijing 100049, China}
\affiliation{Lanzhou University, Lanzhou 730000, China}
\author{Xingbo Zhao}
\email{xbzhao@impcas.ac.cn} 
\affiliation{Institute of Modern Physics, Chinese Academy of Sciences, Lanzhou 730000, China}
\affiliation{School of Nuclear Science and Technology, University of Chinese Academy of Sciences, Beijing 100049, China}
\author{Yang Li}
\email{leeyoung@iastate.edu} 
\affiliation{Department of Physics and Astronomy, Iowa State University, Ames, IA 50011, USA}
\author{Dipankar Chakrabarti}
\email{dipankar@iitk.ac.cn} 
\affiliation{Department of Physics, Indian Institute of Technology Kanpur, Kanpur 208016, India}
\author{James P. Vary}
\email{jvary@iastate.edu} 
\affiliation{Department of Physics and Astronomy, Iowa State University, Ames, IA 50011, USA}

\collaboration{BLFQ Collaboration}

\begin{abstract}
We obtain the electromagnetic form factors, the axial form factor, and the parton distribution functions of the proton from the eigenstates of a light-front effective Hamiltonian in the leading Fock representation suitable for low-momentum scale applications. 
The electromagnetic and the axial form factors are found to be in agreement with the available experimental data. The unpolarized, the helicity, and the transversity valence quark distributions, after QCD evolution, are consistent with the global QCD analyses. The tensor charge also agree the experimental data, while the axial charge is somewhat outside the experimental error bar. 
\end{abstract}
\maketitle
\section{Introduction} 
Electromagnetic form factors (FFs) and parton distribution functions (PDFs) have taught us a great deal about the internal structure of the proton. The Fourier transform of these FFs provides information about spatial distributions of the proton's constituents \cite{Miller:2007uy, Carlson:2007xd}. 
Well-known examples include the charge and magnetization distributions. Another essential tool to investigate hadron structure is deep inelastic scattering (DIS), where individual quarks are resolved. 
One can extract the PDFs from DIS, which encode the distribution of longitudinal momentum and polarization carried by the constituents. With an effective Hamiltonian for constituent quarks, suitable for low-resolution probes, we solve for the proton's light-front wavefunctions (LFWFs) used to produce the FFs and the PDFs. We compare our FFs with experimental data. 
We apply QCD evolution to our initial PDFs in order to incorporate degrees of freedom relevant to higher-resolution probes which allows us to compare our QCD-evolved PDFs with global fits to experimental data.

The matrix element of the electromagnetic current for the nucleon requires two independent FFs:
the Dirac and Pauli FFs. We refer to Refs~\cite{Gao,Hyd,Punjabi,Pacetti:2015iqa,Punjabi:2015bba} for reviews of the experimental results and models on this subject. 
The nucleon electromagnetic FFs
have been theoretically investigated in Refs.~\cite{Alexandrou:2017ypw,Chambers:2017tuf,Alexandrou:2018sjm,Shintani:2018ozy,He:2017viu,Alarcon:2017lhg,Abidin:2009hr,Gutsche:2017lyu,Mondal:2016xpk,Sufian:2016hwn,Brodsky:2014yha,Mondal:2015uha,Chakrabarti:2013dda,Chakrabarti:2013gra,Ye:2017gyb,Gutsche:2014yea,Gutsche:2013zia,Cloet:2012cy,Pasquini:2007iz,Mondal:2016afg}, while their flavor decomposition has been reported in Refs.\cite{Cates:2011pz,Qattan:2012zf,Diehl:2013xca}. Our work is motivated by the advent of high precision measurements of both proton and neutron FFs from ongoing and forthcoming experiments at Jefferson Lab \cite{1,2,3,4,5,6,7}.

PDFs reveal the internal structure of the nucleon in terms of number densities of confined quarks and gluons. At first approximation (``leading twist''), the spin structure of the nucleon is described in terms of three independent PDFs: the unpolarized $f_1(x)$, the helicity $g_1(x)$, and the transversity $h_1(x)$, where $x$ is the light-front longitudinal momentum fraction of the nucleon carried by quarks of flavor $q$. 
We provide these PDFs to assist in the analysis and interpretation of scattering experiments now and in the LHC era.  

While the unpolarized and the helicity PDFs
are fairly well determined~\cite{Ball:2017nwa,Harland-Lang:2014zoa,Dulat:2015mca,Alekhin:2017kpj,Martin:2009iq,Aidala:2012mv,Deur:2018roz,Zheng:2003un,Zheng:2004ce,Parno:2014xzb,Dharmawardane:2006zd,Airapetian:2003ct,Airapetian:2004zf,Alekseev:2010ub,deFlorian:2008mr,deFlorian:2009vb,Nocera:2014gqa,Jimenez-Delgado:2014xza,Ethier:2017zbq,Roberts:2013mja,Liu:2019vsn,Liu:1999ak,Liu:1993cv,Horsley:2012pz,Chambers:2017dov,Radyushkin:2017cyf,Orginos:2017kos,Ji:2013dva,Lin:2014zya,Alexandrou:2015rja,Alexandrou:2016jqi,Ma:2017pxb,Liu:1993cv,Horsley:2012pz,Chambers:2017dov,Radyushkin:2017cyf,Orginos:2017kos,Ji:2013dva,Lin:2014zya,Alexandrou:2015rja,Alexandrou:2016jqi,Ma:2017pxb,Lin:2017snn}, much less information is available on the transversity PDF.
This distribution  is important since it encodes the correlation between the transverse polarization of the constituents and the transverse polarization of the nucleon~\cite{Jaffe:1991kp,Anselmino:2007fs,Anselmino:2008jk,Anselmino:2013vqa,Kang:2015msa,Radici:2015mwa,Radici:2016lam,Bacchetta:2011ip,Bacchetta:2012ty,Radici:2018iag}. 
One of our aims is to predict the transversity PDF, compare with available data, and help motivate future challenging experiments.

Here, we adopt an effective light-front Hamiltonian between quarks and solve for its mass eigenstates at the scale suitable for low-resolution probes with the theoretical framework of basis light-front quantization (BLFQ)~\cite{Vary:2009gt,Zhao:2014xaa,Wiecki:2014ola,Li:2015zda}. Our Hamiltonian includes the holographic QCD confinement potential~\cite{Brodsky:2014yha} supplemented by the longitudinal confinement~\cite{Li:2015zda,Li:2017mlw} along with the one-gluon exchange (OGE) interaction~\cite{Li:2015zda} to account for the dynamical spin effects. By solving this Hamiltonian for the LFWFs in the constituent valence quark Fock space, and fitting the quark mass, confining strength, and coupling constants, we obtain good quality descriptions of its electromagnetic and axial FFs, radii, PDFs, axial and tensor charges. 

\section{Proton wavefunctions from an effective Hamiltonian}
The structures of hadronic bound states are encoded in the LFWFs which are obtained as the eigenfunctions of the LF eigenvalue equation:
$
H_{\mathrm{eff}}\vert \Psi\rangle=M^2\vert \Psi\rangle,
$
where $H_{\mathrm{eff}}$ is the effective Hamiltonian of the system with the mass squared ($M^2$) eigenvalue.  
In general, $\vert \Psi\rangle$ is the eigenvector in the Hilbert space spanned by all Fock sectors. 
At the initial scale where the proton is described by three active quarks, we adopt the LF effective Hamiltonian $H_{\rm eff}$ defined by
\begin{align}\label{hami}
H_{\rm eff}=&\sum_a \frac{{\vec p}_{\perp a}^2+m_{a}^2}{x_a}+\frac{1}{2}\sum_{a\ne b}\kappa^4 \Big[x_ax_b({ \vec r}_{\perp a}-{ \vec r}_{\perp b})^2\nonumber\\
&-\frac{\partial_{x_a}(x_a x_b\partial_{x_b})}{(m_{a}+m_{b})^2}\Big]
+\frac{1}{2}\sum_{a\ne b} \frac{C_F 4\pi \alpha_s}{Q^2_{ab}}\nonumber\\&\times \bar{u}_{s'_a}(k'_a)\gamma^\mu{u}_{s_a}(k_a)\bar{u}_{s'_b}(k'_b)\gamma^\nu{u}_{s_b}(k_b)g_{\mu\nu},
\end{align}
where $\sum_a x_a=1$, and $\sum_a {\vec p}_{\perp a}=0$. $m_{a/b}$ is the mass of the quark, and $\kappa$ is the strength of the confinement. $x_a$ represents the LF momentum fraction carried by quark $a$. Meanwhile, $\vec{p}_\perp$ is the relative transverse momentum, while $\vec{r}_\perp={ \vec r}_{\perp a}-{ \vec r}_{\perp b}$, related to the holographic variable~\cite{Brodsky:2014yha}, is the transverse separation between two quarks. The last term in the effective Hamiltonian represents the OGE interaction where
$Q^2_{ab}=-q^2=-(1/2)(k'_a-k_a)^2-(1/2)(k'_b-k_b)^2$ is the average momentum transfer squared. $C_F =-2/3$ is the color
factor, $g_{\mu\nu}$ is the metric tensor and $\alpha_s$ is the coupling constant. Here, ${u}_{s_a}(k_a)$ is the solution of the Dirac equation, with the subscript $s_a$ representing spin and $k_a$ is the momentum of quark $a$.

Following BLFQ, we expand $\ket{\Psi}$ in terms of the two dimensional harmonic oscillator (`2D-HO') basis in the transverse direction and the discretized plane-wave basis in the longitudinal direction \cite{Vary:2009gt,Zhao:2014xaa}. Each single-quark basis state is identified using four quantum numbers, $\bar \alpha = \{k,n,m,\lambda\}$. The longitudinal momentum of the particle is characterized by the quantum number $k$. The longitudinal coordinate $x^\LCm$ is confined to a box of length $2L$ with anti-periodic boundary conditions for fermions. As a result, the longitudinal momentum $p^\LCp=2\pi k/L$ is discretized, where the dimensionless quantity $k=\frac{1}{2}, \frac{3}{2}, \frac{5}{2}, ...$.
All many-body basis states are selected to have the same total longitudinal momentum $P^+=\sum_ip_i^+$,
where the sum is over the quarks. We rescale $P^+$ using $K=\sum_i k_i$ such that 
$P^+=\frac{2\pi}{L}K$. For a given quark $i$, the longitudinal momentum fraction $x$ is defined as
$
x_i=p_i^+/P^+=k_i/K.
$

The quantum numbers, $n$ and $m$, denote radial excitation and angular momentum projection, respectively, of the particle within the 2D-HO basis, $\phi_{nm}(\vec{p}_\perp)$ \cite{Vary:2009gt,Zhao:2014xaa}. 
The 2D-HO basis should form an efficient basis for systems subject to QCD confinement. For the quark spin, $\lambda$ is used to label the helicity. Our multi-body basis states have fixed values of the total angular momentum projection
$
M_J=\sum_i\left(m_i+\lambda_i\right).
$

The valence wavefunction in momentum space is then expanded as:
\begin{align}
&\Psi^{M_J}_{\{x_i,\vec{p}_{\perp i},\lambda_i\}}=\sum_{\{n_im_i\}}\Bigg\{\psi(\{\bar{\alpha}_i\})\prod_{i=1}^{3}\frac{1}{b}\left(\frac{|\vec{p}_{\perp i}|}{b}\right)^{|m_i|}\nonumber\\
&\sqrt{\frac{4\pi\times n_i!}{(n_i+|m_i|)!}}e^{i m_i\theta_i} L_{n_i}^{|m_i|}\left(-\frac{\vec{p}_{\perp i}^2}{b^2}\right)\exp{\left(-\frac{\vec{p}_{\perp i}^2}{2b^2}\right)}\Bigg\},\label{eq:psi_rs_basis_expansions}
\end{align}
where $\psi(\{\bar{\alpha}_i\})$ is the LFWF in the BLFQ basis obtained by diagnalizing Eq. (\ref{hami}) numerically. $b=0.6$ GeV is the HO scale parameter and $\tan(\theta)=p_2/p_1$. Here $L_n^{|m|}$ is the associated Laguerre function.
We truncate the infinite basis by introducing limit $K_{\rm max}$ such that, $\sum_i k_i = K_{\rm max}$. In the transverse direction, we also truncate by limiting $N_\alpha=\sum_i (2n_i+| m_i |+1)$ for multi-particle basis state to $N_\alpha \le N_{\text{max}}$. 
The basis truncation corresponds to having a UV regulator $\Lambda_{\rm UV}\sim b\sqrt{N_{\rm max}}$. Since we are modeling the proton at a low-resolution scale, we select $N_{\rm max}=10$ and $K_{\rm max}=16.5$. To attempt to simulate the effect of higher Fock spaces and the other QCD interactions,  we use a different quark mass in the kinetic energy, $m_{\rm q/KE}$ and the OGE interaction, $m_{\rm q/OGE}$. We set our parameters $\{m_{\rm q/KE},~m_{\rm q/OGE},~\kappa,~\alpha_s\}=\{0.3~{\rm GeV},~0.2~{\rm GeV},~0.34~{\rm GeV},~1.1\}$ to fit the proton mass and the flavor Dirac FFs \cite{Cates:2011pz,Qattan:2012zf,Diehl:2013xca}. For numerical convenience, we use a small gluon mass regulator ($\mu_g=0.05$ GeV) in the OGE interaction. We find that our results are insensitive to $0.08>\mu_g>0.01$ GeV. 
\section{Form Factors and PDFs of the proton}

In the LF formalism, the flavor Dirac $F^q_1(Q^2)$ and Pauli $F^q_2(Q^2)$ FFs in the proton can be expressed in terms of overlap integrals as \cite{BDH}
\begin{align}\label{eq_DF}
F_1^q(Q^2)=& 
\int_D \Psi^{\uparrow*}_{\{x^{\prime}_i,
\vec{p}^{\prime}_{\perp i},\lambda_i\}} \Psi^{\uparrow}_{\{x_i,
\vec{p}_{\perp i},\lambda_i\}},   \nonumber \\
F_2^q(Q^2)=& -\frac{2 M }{(q^1-iq^2)}
\int_D \Psi^{\uparrow*}_{\{x^{\prime}_i,
\vec{p}^{\prime}_{\perp i},\lambda_i\}} \Psi^{\downarrow}_{\{x_i,
\vec{p}_{\perp i},\lambda_i\}},
\end{align}
with
$\int_D\equiv \sum_{\lambda_i}\int \prod_i[{dx d^2
\vec{p}_{\perp}\over 16 \pi^3}]_i 16 \pi^3 \delta\left(1-\sum x_j\right)
\delta^2\left(\sum \vec{p}_{\perp j}\right).
$
For the struck quark of flavor $q$, 
 ${x'}_1=x_1; 
~{\vec{p}'}_{\perp 1}=\vec{p}_{\perp 1}+(1-x_1) \vec{q}_\perp$ and  ${x'}_i={x_i}; ~{\vec{p}'}_{\perp i}=\vec{p}_{\perp i}-{x_i} \vec{q}_\perp$ for  the spectators ($i=2,3$). We consider the frame where the momentum transfer $q=(0,0,\vec{q}_{\perp})$, thus $Q^2=-q^2=\vec{q}_{\perp}^2$.

Under charge and isospin symmetry, the
proton FFs can be obtained from the flavor FFs \cite{Cates:2011pz}:
$
F_i^{p}=e_u F_i^{u} + e_d F_i^{d},
$
where $e_u(e_d)={2\over3}(-{1\over3})$, with the normalizations $F_1^u(0)=2, F_2^u(0)=\kappa_u$ and $F_1^d(0)=1, F_2^d(0)=\kappa_d$ where the anomalous magnetic moments for the up and the down quarks are $\kappa_u=2\kappa_p+\kappa_n=1.673$ and $\kappa_d=\kappa_p+2\kappa_n=-2.033$.
The nucleon Sachs FFs are again written in terms of Dirac and Pauli FFs as
\begin{align}
 G_E^p(Q^2)&= F_1^p(Q^2)-\frac{Q^2}{4M_p^2}F_2^p(Q^2),\nonumber\\
 G_M^p(Q^2)&= F_1^p(Q^2)+F_2^p(Q^2),
\end{align}
and the electromagnetic radii are defined by
 $
\langle r^2_{E}\rangle=-6 \frac{d G_{E}(Q^2)}{dQ^2}{\Big\vert}_{Q^2=0}
$ and  $
\langle r^2_{M}\rangle=-\frac{6}{G_{M}(0)} \frac{d G_{M}(Q^2)}{dQ^2}{\Big\vert}_{Q^2=0}.
$

\begin{figure}[htp]
\includegraphics[width=0.43\textwidth]{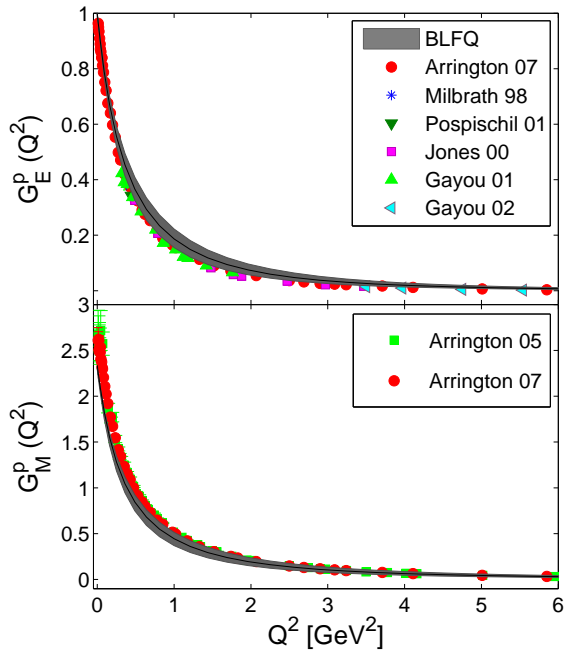}
\caption{Proton Sach's FFs $G_E^p (Q^2)$ (upper panel), and  $G_M^p (Q^2)$ (lower panel)
as functions of $Q^2$. The gray bands are BLFQ results reflecting our $\alpha_s$ uncertainty of $10\%$. The experimental data are taken from Refs.\cite{Gay1,Gay2,Arr,Milbrath,Posp,Jones} and \cite{Arr,Arr2}.
}
\label{figGEM}
\end{figure}

In Fig. \ref{figGEM}, we show the $Q^2$ dependence of the proton electric and the magnetic Sach's FFs. 
Overall, we  obtain a reasonable agreement between theory and experiment for the proton electric FFs. At large $Q^2$, the magnetic form factor is also in good agreement with the data. However, our magnetic form factor at low $Q^2$ exhibits a small deviation from the data. It should be noted that the neglected higher Fock components $|qqqq\bar{q}\rangle$ can have a significant effect on the magnetic form factor \cite{Sufian:2016hwn}.

We present our computed radii in Table \ref{table} and compare with measured data~\cite{Bezginov:2019mdi,pdg} as well as with recent lattice QCD calculations~\cite{Alexandrou:2018sjm}. Here again, we find reasonable agreement with experiment.

\begin{table}[ht]
 \caption{Proton radii, axial and tensor charges, first moments of transversity distributions. Our results are compared with the extracted data and recent lattice QCD calculations. 
 }
\centering 
\begin{tabular}{cccccc}
 \hline\hline
Quantity & BLFQ & Extracted data & Lattice & \\ \hline
%
   $r_E$~fm & $0.802^{+0.042}_{-0.040}$ & $0.833\pm 0.010$ \cite{Bezginov:2019mdi}& $0.742(13)$ \cite{Alexandrou:2018sjm}\\ 
 $r_M$~fm & $0.834^{+0.029}_{-0.029}$ & $0.851\pm 0.026$ \cite{pdg}& $0.710(26)$ \cite{Alexandrou:2018sjm}\\
 $r_A$~fm & $0.680^{+0.070}_{-0.073}$ & $0.667\pm 0.12$ \cite{Hill:2017wgb}& $0.512(34)$ \cite{Yao:2017fym}\\
 \hline
 $g_A$ & $1.41^{+0.06}_{-0.06}$ & $1.2723\pm 0.0023$ \cite{pdg}& $1.237(74)$ \cite{Yao:2017fym}\\
 \hline
 $g_T^d$ & $-0.20^{+0.02}_{-0.04}$ & $-0.25^{+0.30}_{-0.10}$ \cite{Anselmino:2013vqa}& $ -0.204(11)$ \cite{Gupta:2018lvp}\\
 $g_T^u$ & $0.94^{+0.06}_{-0.15}$ & $0.39^{+0.18}_{-0.12}$ \cite{Anselmino:2013vqa}& $0.784(28)$ \cite{Gupta:2018lvp}\\
  $\langle x\rangle_T^ {u-d}$ & $0.229^{+0.019}_{-0.048}$ & -- & $0.203(24)$ \cite{Alexandrou:2019ali}\\
\hline\hline
 \end{tabular}
\label{table} 
\end{table}

\begin{figure}[htp]
\includegraphics[width=0.43\textwidth]{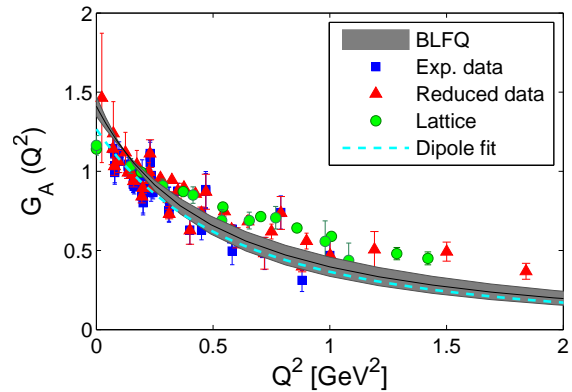}
\caption{Axial-vector form factor $G_A=G_A^{u}-G_A^{d}$ as function of $Q^2$. The gray band is the BLFQ result. The extracted data are taken from the Refs. \cite{Bernard:2001rs,Hashamipour:2019pgy} and the lattice data from \cite{Alexandrou:2013joa}. The dashed line represents the dipole fit of the experimental data \cite{Bernard:2001rs}.
}
\label{figGa}
\end{figure}

The axial form factor, which is identified with the matrix elements of axial-vector local operator, can also be expressed in terms of LFWFs
\begin{align}\label{eq_AF}
G_A^q(Q^2)=
\int_D (\Lambda)~\Psi^{\uparrow*}_{\{x^\prime_i,
\vec{p}^\prime_{\perp i},\lambda_i\}} \Psi^{\uparrow}_{\{x_i,
\vec{p}_{\perp i},\lambda_i\}}. 
\end{align}
Here, $\Lambda=1(-1)$ depends on the struck quark helicity $\lambda_1=\frac{1}{2}(-\frac{1}{2})$. Experimental information about the axial FFs is very limited. Until now, there are only two sets of experiments: (anti)neutrino scattering off protons or nuclei and charged pion electroproduction \cite{Bernard:2001rs,Schindler:2006jq}. Fig. \ref{figGa} shows the results obtained for the axial form
factor, $G_A=G_A^u-G_A^d$ as a function of $Q^2$, where we compare our BLFQ results with the experimental data~\cite{Bernard:2001rs,Hashamipour:2019pgy} and with lattice results~\cite{Alexandrou:2013joa}.
Considering the theoretical and experimental uncertainties, the agreement is good. At $Q^2=0$, the axial form factor is the axial charge, $g_A=G_A(0)$. $g_A$ is quoted in Table~\ref{table}, where we see our value differs somewhat from extracted data~\cite{pdg}  and lattice \cite{Yao:2017fym}.   
We also evaluate the axial radius from:
 $
\langle r^2_{A}\rangle=\frac{6}{g_A} \frac{d G_{A}(Q^2)}{dQ^2}{\Big\vert}_{Q^2=0}
$. 
As can be seen from the Table \ref{table}, the BLFQ result is in good agreement with the extracted data from the analysis of neutrino-nucleon scattering experiment \cite{Hill:2017wgb,Meyer:2016oeg}.

With our LFWFs, the proton's valence quark PDFs at leading twist are given by
\begin{align}\label{eq_PDF}
f_1^q=&
\int_D \Psi^{\uparrow*}_{\{x^\prime_i,
\vec{p}^\prime_{\perp i},\lambda_i\}} \Psi^{\uparrow}_{\{x_i,
\vec{p}_{\perp i},\lambda_i\}} \delta(x-x_1), \nonumber
\\
g_1^q=&
\int_D (\Lambda)~\Psi^{\uparrow*}_{\{x^\prime_i,
\vec{p}^\prime_{\perp i},\lambda_i\}} \Psi^{\uparrow}_{\{x_i,
\vec{p}_{\perp i},\lambda_i\}}\delta(x-x_1) , 
\\
h_1^q=&
\int_D \Big[\Psi^{\uparrow*}_{\{x^\prime_i,
\vec{p}^\prime_{\perp i},\lambda^\prime_i\}} \Psi^{\downarrow}_{\{x_i,
\vec{p}_{\perp i},\lambda_i\}}+(\uparrow \leftrightarrow \downarrow)\Big]\delta(x-x_1) , \nonumber
\end{align}
where the $\lambda^\prime_1=-\lambda_1$ and $\lambda^\prime_{2,3}=\lambda_{2,3}$.
 At the model scale relevant to constituent quark masses which are several hundred $\mathrm{MeV}$, the unpolarized PDFs for the valence quarks are normalized as
$
\int_{0}^{1}f_1^u(x)\,dx =2, \int_{0}^{1}f_1^d(x)\,dx=1.
$
We also have the following momentum sum rule: 
$
\int_{0}^{1}x\,f_1^u(x)\,dx +\int_{0}^{1}x\,f_1^d(x)\,dx=1, 
$

Next, to evolve our PDFs from our model scale, defined as $\mu_{0}^2$, to a higher scale $\mu^2$,
we adopt the next-to-next-to-leading order (NNLO) Dokshitzer-Gribov-Lipatov-Altarelli-Parisi (DGLAP) equations~\cite{Dokshitzer:1977sg,Gribov:1972ri,Altarelli:1977zs} of QCD.  This scale evolution allows quarks to emit and absorb gluons, with the emitted gluons capable of producing quark-antiquark pairs as well as additional gluons. In this picture, the sea quark and gluon components of the constituent quarks are revealed at the higher scale through QCD. 
While applying the DGLAP equations numerically~\cite{Salam:2008qg}, we impose the condition that the running coupling $\alpha_s(\mu^2)$ saturates in the infrared at a cutoff value of max $\alpha_{s}=1$ \cite{Lan:2019vui,Lan:2019rba}, consistent with our fit value discussed above.
We determine ${\mu_{0}^2=0.195\pm0.020~\rm{GeV}^2}$ by requiring the result after QCD evolution to produce the total first moments of the valence quark unpolarized PDFs from the global data fits with average values, $\langle x\rangle_{u_v+d_v}=0.37\pm 0.01$ at $\mu^2=10$ GeV$^2$~\cite{deTeramond:2018ecg}.

\begin{figure}[htp]
\includegraphics[width=0.43\textwidth]{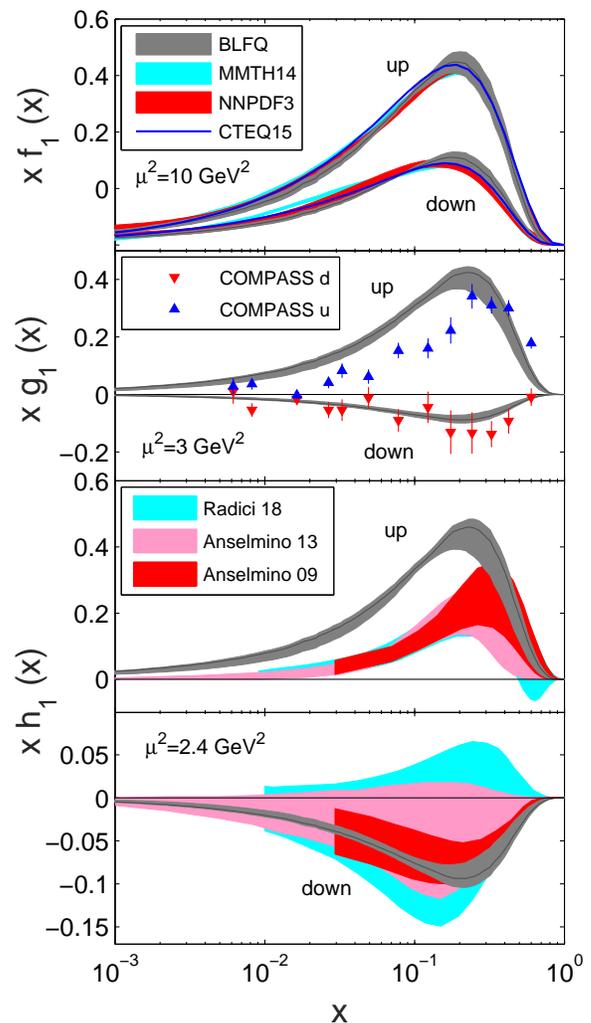}
\caption{Top panel: comparison for $xf_1(x)$ in the proton from BLFQ (gray
bands) and global fits: MMHT14~\cite{Harland-Lang:2014zoa} (cyan bands), NNPDF3.0 \cite{Ball:2017nwa} (red bands), and CTEQ15 \cite{Dulat:2015mca} (blue line). Second panel: comparison for $xg_1(x)$ in the proton from BLFQ (black
bands) and measured data from COMPASS \cite{Alekseev:2010ub}. Lower two panel: Comparison for $xh_1(x)$ in the proton from BLFQ (gray
bands) and global analyses by Radici {\it et. al.} \cite{Radici:2018iag} (cyan bands) and  Anselmino {\it et. al.} \cite{Anselmino:2013vqa} (pink bands), and \cite{Anse09} (red bands).
}
\label{figfg}
\end{figure}

Fig.~\ref{figfg} shows our results for the valence quark unpolarized and spin dependent PDFs of the proton, where we compare the valence quark distribution after QCD evolution with the  global fits by MMHT14~\cite{Harland-Lang:2014zoa}, NNPDF3.0 \cite{Ball:2017nwa} and CTEQ15 \cite{Dulat:2015mca} Collaborations. The error bands in our evolved distributions are due to the spread in the initial scale $\mu_{0}^2=0.195\pm 0.020$ GeV$^2$ and the uncertainties in the coupling constant, $\alpha_s=1.1\pm0.1$. Our unpolarized valence PDFs for both up and down quarks are found to be in good agreement with the global fits. Meanwhile, we evolve the spin dependent PDFs from our model scale to the relevant experimental scale $\mu^2=3$ GeV$^2$ and find that the down quark helicity PDF agrees well with measured data from COMPASS Collaboration \cite{Alekseev:2010ub}. However, for the up quark, our helicity PDF tends to overestimate the data below $x\sim 0.3$. 

The transversity distribution at $\mu^2=2.4$ GeV$^2$ is also shown in Fig.~\ref{figfg}. We compare our prediction with the global analysis of pion-pair production in DIS and in proton-proton
collisions with a transversely polarized proton by Radici {\it et.~al.} \cite{Radici:2018iag}, the global analysis of the data on azimuthal asymmetries in SIDIS, from the HERMES and COMPASS Collaborations, and  $e^+e^-$ data from the BELLE Collaboration by Anselmino {\it et.~al.} \cite{Anse09,Anselmino:2013vqa}. 
Our down quark transversity distribution is in accord with the fits. The up quark distribution in our approach deviates in the low $x$ region, while it shows resonable agreement at large $x$ with the global fits. 

Transversity has recently received increasing attention because of the importance of a precise determination of its integral, the so-called
tensor charge $g_T$.
We compare our results at $\mu^2=2.4$ GeV$^2$ with extracted data as well as with lattice data in Table~\ref{table}.
Again we observe that BLFQ predicts the tensor charges quite well for down quark in comparison with the global QCD analysis \cite{Anselmino:2013vqa}. However, for the up quark it deviates from the extracted data but our value is closer to recent lattice data \cite{Gupta:2018lvp} and  our approach yields comparable agreement with results from phenomenological models as well as other lattice calculations~\cite{Gock05,Cloet08,Pasq05,Waka07}. We also provide the first moments of transversity distribution, $\langle x \rangle_T^{u-d}$ in the Table~\ref{table}, which agree reasonably well with lattice data \cite{Alexandrou:2019ali}.
\section{Conclusions}

We present a model for the proton that provides observables from the low resolution constituent quark scale to high resolution experiments. Specifically, we begin with an effective LF Hamiltonian incorporating confinement and one gluon exchange interaction for the valence quarks suitable for low-resolution properties. 
Using basis LF quantization, the LFWFs obtained as the eigenvectors of this Hamiltonian were then used to generate the proton electromagnetic and axial form factors and the initial PDFs for different quark polarizations. 
We have obtained reasonable agreement with the experimental data for the Sach's FFs, the axial form factor as well as the electromagnetic radii for the proton.
The unpolarized, helicity and transversity PDFs at higher scale relevant to global QCD analyses have been computed based on the NNLO DGLAP equations. The initial low-resolution scale is the only adjustable parameter involved in QCD scale evolution and we obtain it by fitting the first moments of unpolarized PDFs from global QCD analyses. We then find the unpolarized, the helicity, and transversity PDFs agree with results from the corresponding global fits or experimental data in Refs.~\cite{Harland-Lang:2014zoa,Ball:2017nwa,Dulat:2015mca}, Ref. \cite{Alekseev:2010ub}, and Refs. \cite{Anse09, Radici:2018iag}, respectively.  The axial charge and the tensor charge also show reasonable agreement with the extracted data or the lattice results. It should also be noted that basis truncation may play a role that should be examined in future research.
The effective LFWFs can be used to study other parton distributions, such as the generalized parton distributions, the transverse momentum dependent
parton distributions and the Wigner distributions. The presented results affirm the utility of our model and motivate application of analogous
effective Hamiltonians to the other hadrons.

\begin{acknowledgments}
We thank Henry Lamm, Wei Zhu, Shuai Liu for many useful discussions. CM is supported by the National Natural Science Foundation of China (NSFC) under the Grant No. 11850410436. XZ is supported by new faculty startup funding by the Institute of Modern Physics, Chinese Academy of Sciences and by Key Research Program of Frontier Sciences, CAS, Grant No ZDBS-LY-7020. JPV is supported by the Department of Energy under Grants No.~DE-FG02-87ER40371, and No.~DE-SC0018223 (SciDAC4/NUCLEI). A portion of the computational resources were provided by the National Energy Research Scientific Computing Center (NERSC), which is supported by the Office of Science of the U.S. Department of Energy under Contract No. DE- AC02-05CH11231.
\end{acknowledgments}

\end{document}